\documentclass[journal]{vgtc}                     

\onlineid{1611}

\vgtccategory{Research}

\usepackage{enumitem}
\usepackage{fontawesome}
\usepackage{colortbl}
\usepackage{url}
\usepackage{float}
\usepackage{pifont}
\usepackage{graphicx}
\usepackage{xurl}
\usepackage{booktabs}
\usepackage{multirow}
\usepackage{makecell} 

\newcommand{\pheading}[1]{\vspace{2px}\noindent\textbf{#1}}
\newcolumntype{R}[1]{>{\raggedright\arraybackslash}p{#1}}
\newcommand{\tabpercent}[1]{\textcolor{black!80}{\scriptsize #1\%}}

\arrayrulecolor{gray} 

\newenvironment{tightItemize}{\begin{itemize} \itemsep
-3pt}{\end{itemize}}

\newenvironment{tightEnumerate}{\begin{enumerate} \itemsep
-3pt}{\end{enumerate}}

\newcommand{\titles}{\textit{Titles}\xspace}
\newcommand{\subtitles}{\textit{Subtitles}\xspace}
\newcommand{\annotations}{\textit{Annotations}\xspace}
\newcommand{\captions}{\textit{Captions}\xspace}
\newcommand{\axes}{\textit{Axes}\xspace}
\newcommand{\legends}{\textit{Legends}\xspace}
\newcommand{\paragraphs}{\textit{Paragraphs}\xspace}
\newcommand{\titleType}{\textit{Title}\xspace}
\newcommand{\subtitle}{\textit{Subtitle}\xspace}
\newcommand{\annotationType}{\textit{Annotation}\xspace}
\newcommand{\captionType}{\textit{Caption}\xspace}
\newcommand{\axis}{\textit{Axis}\xspace}
\newcommand{\legend}{\textit{Legend}\xspace}

\newcommand{\identifyMappings}{\textsc{Identify Mappings}\xspace}
\newcommand{\identifyValues}{\textsc{Identify Values}\xspace}
\newcommand{\presentMetadata}{\textsc{Present Metadata}\xspace}
\newcommand{\replaceMappings}{\textsc{Replace Mappings}\xspace}
\newcommand{\compareMappings}{\textsc{Compare Mappings}\xspace}
\newcommand{\compareValues}{\textsc{Compare Values}\xspace}
\newcommand{\summarizeValues}{\textsc{Summarize Values}\xspace}
\newcommand{\summarizeConcepts}{\textsc{Summarize Concepts}\xspace}
\newcommand{\summarizeConceptsVar}{\textsc{Summarize Concepts: Variables}\xspace}
\newcommand{\summarizeConceptsSyn}{\textsc{Summarize Concepts: Synthesis}\xspace}
\newcommand{\presentContext}{\textsc{Present Context}\xspace}
\newcommand{\presentValence}{\textsc{Present Valenced Subtext}\xspace}

\newcommand{\fOne}{\texttt{F1}\xspace}

\newcommand{\attribution}{\texttt{Attribution and Variables}\xspace}
\newcommand{\fTwo}{\texttt{F2}\xspace}
\newcommand{\fTwoAnnotation}{\texttt{F2:Annotation-Centric Design}\xspace}
\newcommand{\annotationFactor}{\texttt{Annotation-Centric Design}\xspace}
\newcommand{\fThree}{\texttt{F3}\xspace}

\newcommand{\visual}{\texttt{Visual Embellishments}\xspace}
\newcommand{\fFour}{\texttt{F4}\xspace}
\newcommand{\fFourNarrative}{\texttt{F4:Narrative Framing}\xspace}
\newcommand{\narrative}{\texttt{Narrative Framing}\xspace}

\title{An Analysis of Text Functions in Information Visualization}

\author{%
  \authororcid{Chase Stokes}{0000-0001-7644-9021},
  \authororcid{Anjana Arunkumar}{0000-0003-3513-8600},
  \authororcid{Marti A. Hearst}{0000-0002-4346-1603},
  and
  \authororcid{Lace Padilla}{0000-0001-9251-5279}
}

\authorfooter{
  \item
  	Chase Stokes and Marti Hearst are with University of California, Berkeley.
  	E-mail: chase\_stokes, hearst@berkeley.edu
  \item
  	Anjana Arunkumar and Lace Padilla are with Northeastern University.
        E-mail: a.arunkumar, l.padilla@northeastern.edu
}

\abstract{%
Text is an integral but understudied component of visualization design. Although recent studies have examined how text elements (e.g., titles and annotations) influence comprehension, preferences, and predictions, many questions remain about textual design and use in practice. This paper introduces a framework for understanding text functions in information visualizations, building on and filling gaps in prior classifications and taxonomies. Through an analysis of 120 real-world visualizations and 804 text elements, we identified ten distinct text functions, ranging from identifying data mappings to presenting valenced subtext. We further identify patterns in text usage and conduct a factor analysis, revealing four overarching text-informed design strategies: Attribution and Variables, Annotation-Centric Design, Visual Embellishments, and Narrative Framing. In addition to these factors, we explore features of title rhetoric and text multifunctionality, while also uncovering previously unexamined text functions, such as text replacing visual elements. Our findings highlight the flexibility of text, demonstrating how different text elements in a given design can combine to communicate, synthesize, and frame visual information. This framework adds important nuance and detail to existing frameworks that analyze the diverse roles of text in visualization. 
}

\keywords{Visualization, text, language, text function, factor analysis, design patterns.}

\teaser{
  \centering
  \includegraphics[width=\linewidth, alt={xxxx}]{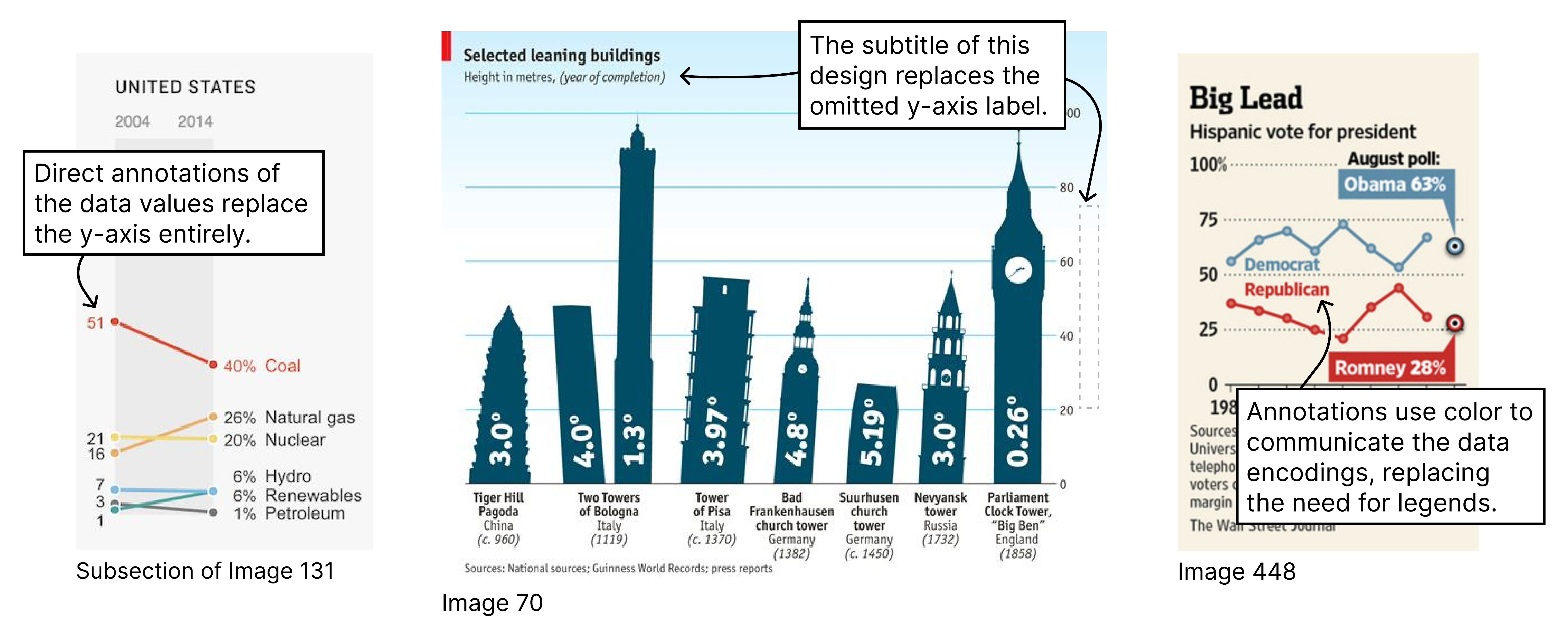}
  \caption{
 These designs demonstrate \replaceMappings: the use of text elements to replace conventional mapping components in a chart, such as \axes and \legends. 
  This function showcases the integration of text and visual design choices.
  Within this paper, we propose and analyze ten distinct text functions, including \replaceMappings, and show how they can be organized into four factors.
  }
  \label{fig:teaser}
}

\graphicspath{{figs/}{figures/}{pictures/}{images/}{./}} 

\usepackage{tabu}                      
\usepackage{booktabs}                  
\usepackage{lipsum}                    
\usepackage{mwe}                       

\usepackage{mathptmx}                  

\begin{document}

\maketitle


\section{Introduction}

Visualization research has produced important empirical and theoretical work that provides recommendations for using graphical elements (e.g., \cite{ware2019information, card1999using, Bertin1983semiology, tufte1985visual}) while historically overlooking the impact of text. Designers use text in almost every visualization, but unlike visual design standards that optimize data encodings for precision and clarity, text-related decisions often rely on designer intuition and conventions.
Text plays a critical role in visualizations; titles, captions, and annotations provide context, guide attention and frame understanding. Language is a highly flexible tool, offering countless ways to convey an idea, making it a promising subject for deep analysis.

Recent studies have begun to reveal how variations in text placement, function, and style influence reader experience and comprehension (e.g., \cite{stokes2024delays, stokes2022more, stokes2023role, stokes2022striking, zhu2022captions, kim2021towards, ajani2021declutter}), and    prior work has built useful foundations for  visualization-based text classifications \cite{lundgard2021accessible, rahman2024qualitative, liu2023autotitle, brehmer2013multi, amar2004knowledge}.
However, there remain many open questions, including a gap in understanding the specific functions that textual elements perform. 
Previous efforts often focus on broad categories or isolated uses of text and may not clearly define the specific functions of text elements. 
The lack of a clear, functional framework creates a barrier to understanding how text works in concert with visual elements to support data communication.

We define \textbf{text \textit{function} as the specific role a text element plays in supporting the design and data communication of a visualization, focusing on the actions it performs and the target it operates on.}
The ``action'' terms borrow verbs directly from prior taxonomies in visualizations \cite{rahman2024qualitative, brehmer2013multi}. The ``target'' terms were compiled through the exploratory analysis described in \cref{sec:constructing_functions} as well as close readings of taxonomic definition in prior work \cite{rahman2024qualitative, lundgard2021accessible}.
A \textit{function} is distinct from the broader notion of the role of text, as studied in prior works \cite{stokes2023role, stokes2022striking, rahman2024qualitative}, which describes the general purpose or effect of text in guiding reader interpretations.

In an effort to further clarify this distinction, we investigated our first research question: \textbf{RQ1: What are the \textit{functions} of text in visualization designs?} 
This question seeks to identify and characterize the detailed functions individual text components fulfill.
We develop this text function framework via a detailed qualitative analysis of 804 individual text components drawn from 120 visualizations from several domains, such as blogs, government reports, and news sources.
In our framework, functions describe how visualizations use language at the level of an individual text component, analyzed independently from but in conjunction with its type (e.g., \titles, \legends, \annotations).

Through the qualitative analysis, we uncover several novel insights into the roles text plays in visualization design. For instance, we find that text usage in visualizations is often multifunctional; \titles can state the chart's variables, synthesize data, compare values, and hint at emotional valence all at once. Another insight is shown in \cref{fig:teaser}.
\replaceMappings is a previously unexamined function in which a text element replaces a conventional component of a visualization, as when data values on the chart replace the y-axis.

The detailed understanding of text function can be used to create stable characterizations of text in visualization design. 
Based on the foundational understanding of text functions distilled in \textbf{RQ1}, we explore \textbf{RQ2: What text design patterns emerge across visualizations, and which functions are engaged in each pattern?}

We conducted a factor analysis to investigate combinations of text functions within text elements and entire designs, aiming to uncover common design patterns.
This analysis revealed four characterizing factors of how text is used across designs, which we named: \attribution, \annotationFactor, \visual, and \narrative. 
These factors not only highlight trends in text usage but also help clarify how specific text choices can align with broader design goals.

\pheading{Contributions:} 
In this work, we build on existing frameworks for interpreting text in visualizations, focusing on the function of each text component with respect to its content and how it interacts with other elements in the design. 
We contribute: 

\begin{itemize}
    \item \textbf{A Set of Text Functions}: We develop a structured set of text functions, based on prior work and our own detailed analysis. These functions fill existing gaps in text taxonomies.
    \item \textbf{Application of Text Functions}: We apply these function definitions to more than 800 text components using a detailed codebook, achieving high inter-annotator agreement ($Mean = 0.97$). We show how text functions occur across individual text types (e.g., titles), how often a single text element exhibits multiple functions, and other notable uses distinct from overarching design patterns.
    \item \textbf{Design Patterns Based on Text Function}: We examine how text functions are used across text components within a given design by performing a factor analysis on the coded data. We extract four factors that represent the main trends in text design across the corpus. We conduct computational validation of these factors to ensure the robustness and stability of item associations.
\end{itemize} 

Clearer guidance on text's varied roles may help designers use language more effectively. 
By establishing a structured framework for text functions, we provide a foundation for \textit{future} improvements to both human and automated design decisions.

\section{Related Work}

This paper proposes a new framework for classifying the text used in information visualizations.  This section summarizes related work on text in visualization, prior work on taxonomies in the information visualization space, and remaining open questions.

\subsection{Text in Visualizations}

A growing body of work focuses on the combination of text and visual information. 
Text elements of a visualization have a substantial role in directing and attracting attention when viewing a visualization.
Text paragraphs, titles, legends, labels, and axes are fixated upon within the first three seconds of viewing a chart, typically before the reader examines the data displayed \cite{bylinskii2015eye}. 
The content viewers recall from visualizations is often based on information or content they extract from the title \cite{borkin2015beyond}.
Different titles lead to different information recalled from the same visualizations \cite{kong2018frames, kong2019trust}.

Annotations and captions also affect conclusions drawn from visualizations, depending on the content. 
When specific features are highlighted in captions, readers are more likely to mention those features in their takeaways, particularly if they align with visually salient elements \cite{kim2017explaining, zhu2022captions}. 
Annotations similarly guide interpretation, with participants more likely to conclude statistical or external information if it is explicitly written in an annotation \cite{stokes2022striking}.

Text can affect interpretations beyond conclusions; using annotations to highlight one outcome over another for a prediction task led to higher ratings of designer bias compared to neutral text conditions \cite{stokes2023role}.
It is likely that text elements have differing impacts depending on the task or context. In the same study, text had only a minimal effect on the prediction task itself, despite influencing other perceptions of the chart. 

Visualizations that use text to `focus' the view of the data tend to be perceived by readers as clearer and more visually appealing \cite{ajani2021declutter, knaflic2015storytelling}.
Assessment of probability representations revealed that important information was easier to identify with visuals but easier to comprehend with text \cite{ottley2019curious}. 
The impact of multimedia communication may also depend in part on the audience. 
In one study examining decision-making under uncertainty, participants with lower working memory capacity benefited more from the combination of text and visual information than those with a higher working memory capacity \cite{bancilhon2023combining}. 

Research endeavors have primarily focused on the impact of text elements on perceptions and understanding of visualized data, but there are also many open questions about designer \textit{practice} regarding text elements. 
Prior work examining the role of text in interactive dashboards \cite{sultanum2024instruction} demonstrates how text serves both navigational and interpretative roles.
This work additionally considered the use of text formatting (e.g., italics) as a tool for designers to distinguish between different text elements or purposes.
Text and natural language are increasingly incorporated into design tools,
including question-answering pipelines \cite{dhamdhere2017analyza, kim2020answering}, automatically generated summaries \cite{kanthara2022chart, mittal1998describing, tableausummary}, and assistance in data storytelling \cite{metoyer2018coupling, sultanum2023datatales}. 

Visualization researchers have demonstrated a growing interest in leveraging text effectively in visualization design. 
However, the concept of a text's role or function across these studies is often ambiguously defined and context- or task-specific.  
We need to develop a structured and thorough understanding of text function in visualizations to develop better systems and guidelines for designers.
Initial efforts in this area include several variations on taxonomies of text.

\subsection{Visualization Taxonomies}

\subsubsection{Taxonomies of Visualization Tasks}

Scholars have developed structured frameworks 
to help designers navigate the multitude of choices available in visual design. 
Taxonomies in visualization research categorize the elementary tasks that readers complete with charts \cite{amar2005low, lee2006task, shneiderman2003eyes}, high-level comprehension tasks \cite{amar2004knowledge, tory2004rethinking, burns2020evaluate}, abstract combinations of these tasks \cite{brehmer2013multi}.
These frameworks help designers make informed choices on key aspects such as visual encodings, interactions, and spatial organization.

Low-level task taxonomies \cite{amar2005low, lee2006task, shneiderman2003eyes} guide designers on how visualizations might support analytical activities. These taxonomies break down visualization tasks into core components, including ``filter,'' ``sort,'' and ``compare,'' enabling designers to tailor their visualizations to specific analytic needs. 
The tasks have also been evaluated or examined according to different charts or data types, providing more specificity in how different design choices can support different tasks.

In contrast, high-level task taxonomies are typically more focused on the broader contexts of design or use. 
For example, Tory and M{\"o}ller define a taxonomy based on the design models of visualization algorithms, distinguishing between continuous and discrete assumptions about data representation \cite{tory2004rethinking}. 
Burns et al. apply Bloom's taxonomy to visualizations to map and understand different levels of user understanding \cite{burns2020evaluate}. This application of the educational taxonomy has also been used to measure graphical literacy \cite{arneson2018visual, peng2022evaluating}. 

The abstract task taxonomy proposed by Brehmer and Munzner offers a bridge between low- and high-level task classifications by introducing a multi-level abstract framework structured around the questions of why, how, and what tasks are performed \cite{brehmer2013multi}. 
This work suggests a set of verbs that describe the perspective and goals of users (e.g., discover), search actions they may perform (e.g., lookup), and elementary tasks they could perform with the data points (e.g., identify). This encompasses the \textit{why}. The \textit{what} refers to the possible inputs and outputs of the particular task, such as values, structures, or other visualization features. \textit{How} refers to specific actions or interactions from the user (e.g., select, filter). 

\subsubsection{Taxonomies of Text In Visualization}

Emerging research has sought to develop frameworks and taxonomies for textual elements in visualizations. 
Rahman et al. \cite{rahman2024qualitative} built on the verbs from the abstract task taxonomy (identify, compare, summarize, and present) \cite{brehmer2013multi}, conducting an extensive thematic coding of real-world annotations (both visual and textual). This work presents a design space of annotation types, which includes enclosure, connector, text, glyph, color, indicator, and geometric. 
The visual elements could be combined with text information as part of a single annotation grouping.

Using a ``how, why, and what'' structure, Rahman et al. outline a design space for the analytic purposes of the annotations, the strategies available, and the types of data needed to generate the annotations. 
Although this work incorporates text annotations, it does not account for specific communicative functions performed by text; we build on the verb structure from this taxonomy, and others \cite{brehmer2013multi} to explicitly assess the functionality of text.

Liu et al. \cite{liu2023autotitle} developed a structured design space for visualization titles based on a survey of existing practices. Their framework identified two key dimensions: generic information (i.e., variables and encoding schemes) and data features (i.e.,  mathematical operations such as trends and aggregations). 
For example, a title might simply state the variables shown (``Staff Counts Across Healthcare Systems'') or describe a specific trend in the data (``NHS Has Fewer Staff Than Some Counterparts''). 
Their work focused specifically on titles that could be generated automatically from visualization inputs, excluding external context (e.g., current events).  
We use this taxonomy to inform how we define generic and descriptive functionality.

Moving beyond single chart images, Hao et al. examined taxonomies of data-driven news articles \cite{hao2024design}, including how text appeared in visualizations. 
Titles often mirrored headlines, stating issues or emphasizing data features; captions tended to cite sources and scope, occasionally describing the visual content. Annotations typically labeled data values but sometimes conveyed thresholds or summaries.

Prior work on dashboard design \cite{srinivasan2024zoo, bach2023dashboard, sarikaya2019dashboards, sultanum2024instruction} also incorporates studies of how text `blocks' are used to structure and support visualizations. Text in dashboards is often added to provide additional information beyond basic data facts. Rather than focus on visualization structure, our work examines the content \textit{inside} the text blocks themselves.

A related but distinct line of work focuses on the use of text to improve accessibility, particularly through alternative (alt) text. 
To support alt text generation, Lundgard and Satyanarayan proposed a conceptual model with four semantic levels: \textit{encoded} (e.g., chart type), \textit{statistical} (e.g., extrema), \textit{perceptual} (e.g., pattern synthesis), and \textit{contextual} (e.g., current events). 
This model primarily provides guidance for writing alt text but has also been used for evaluating text content in and around visualizations \cite{stokes2022striking, stokes2023role, sultanum2024instruction, zhu2022captions}.

\subsubsection{Gaps in Existing Taxonomies}

After careful review of prior work, we identify several important gaps in the existing formalizations with respect to text in visualization: 

\begin{tightEnumerate}
    \item Text is characterized only at a broad level.
    \item One kind of text use is considered (e.g., titles, alt text), rather than classifying across all uses within a visualization.
    \item A given text element is assigned only a single function.
    \item Interactions between text elements in a design are not considered.
\end{tightEnumerate}

If these gaps remain unaddressed, designers and researchers risk relying on incomplete or oversimplified models that fail to capture the complexity of text use in visualizations, leading to suboptimal design decisions and understanding.
To address these issues, we pose two guiding research questions.
First, RQ1: \textbf{what are the \textit{functions} of text in visualization designs?} This question addresses Gaps 1-3 by clarifying the \textit{specific} roles text can play across a range of elements, including cases where a single element serves multiple functions. 
Building on existing frameworks, we propose a detailed set of text functions. 

We use this set of functions to then investigate RQ2: 
\textbf{what text design patterns emerge across visualizations, and which functions are engaged in each pattern?} 
Addressing Gap 4, this question considers how features of text elements, including their functions, interact and combine within a given design. 
Using a factor analysis on a set of text design variables to systematically group these functions, we provide a thorough assessment of 
the text design space, including possible design goals.
Together, these questions establish a foundation for a structured and nuanced representation of text in visualization.

\section{Methodology for Function Framework Creation}
\label{sec:constructing_functions}

\subsection{Corpus Creation}

We used two collections of visualizations: the first for open coding of functions and the second to analyze the resulting set of functions\footnote{Three images overlapped between the two collections.}.
We examined a subset of real-world visualizations sourced from the Image-to-Information corpus \cite{arunkumar2023image}, which itself draws stimuli from MASSVIS \cite{borkin2015beyond, borkin:2013}.
The original Image-to-Information corpus consisted of 500 images curated by sampling from various sources equivalently and supplementing collected infographics
with more standard visualization types.
We selected this dataset since it represents a broad spectrum of ``in the wild'' visualization designs from diverse sources such as news media, government reports, and scientific literature.

For this paper, we examined a subset (n = 120) of the original corpus, focusing on visualizations that conveyed information through relatively common chart types and structures.
To do so, we excluded a total of 380 designs. First, we filtered out 273 designs classified as ``image'' rather than ``information'' in the original study~\cite{arunkumar2023image}. 
This filtering ensured that our focus remained on visualizations meant to inform readers.

We further refined this subcorpus by removing infographics (54), diagrams (25), tables (10), and interactive designs (2). 
Additionally, we omitted any designs with unreadable text due to poor image resolution. 
These criteria resulted in a set of 120 visualizations, each offering rich text examples for analysis.
Compared to the original Image-to-Information corpus, our subset leans more heavily on visualizations from news media (\textit{n} = 46) and government reports (19).
While our corpus focuses on a narrower range of sources, it captures key visualizations aimed at general audiences, and our text function methodology can be extended to genre-specific studies, as discussed in \cref{sec:fw}.

\subsection{Code Development}

With insights from our review of prior taxonomies, we conducted open coding of 18 existing designs, including designs encountered in the real world and drawn from MASSVIS and other chart corpora \cite{borkin:2013, rahman2024qualitative, lundgard2021accessible}. 
Because the open coding was conducted on a small subset of designs in preparation for closed coding on the primary corpus, we drew from diverse sources to avoid missing potential text functions.

This open coding took place at the sentence or phrase level and consisted of writing short phrases to describe what functions the text was performing (e.g., highlighting an outlier in the data). We then performed axial coding, grouping together similar phrases to distill sets of candidate functions. We continued to review and discuss charts containing text that did not clearly fit within our initial function set, refining our categories accordingly.
Through repeated cycles of open and axial coding, we developed a structured set of distinct text functions to capture the range of ways text contributes to visualization design. 

Building on these coded insights, we returned to foundational taxonomies, such as the abstract task \cite{brehmer2013multi} and annotation taxonomies \cite{rahman2024qualitative}, to further refine the functions. 
From these frameworks, we adopted a verb-focused specification, borrowing the specific verbs (e.g., ``Compare'') used in both taxonomies. 

\subsection{Function Coding}
\label{sec:applying_functions}

We coded the corpus of 120 selected visualizations and 804 distinct text components according to the set of text functions. We manually extracted text from each visualization\footnote{ In initial tests, large-language models (LLMs) extracted text fairly well with some errors (e.g., capturing axes). At the time of analysis, LLMs did not support automatic coding of text with our labels, so we did this work manually.}, recording either the exact content or summarizing repetitive elements such as categorical headers or date labels. For example, we used ``set of categories'' to describe repeated labels on a categorical axis.
Multi-line text was treated as a single element to preserve context. 

We coded additional features of each text element, as shown in \cref{tab:text_metadata}. These codes drew from prior definitions when possible \cite{sultanum2024instruction} and were derived via open coding and discussion. Specific definitions are available in supplementary materials. 
A second author reviewed all extracted text and independently coded this metadata, with discrepancies resolved through discussion. 

\begin{table}[ht!]
    \centering
    \def\arraystretch{1.2}
    \begin{tabular}{|R{{0.16}\linewidth}|R{{0.35}\linewidth}|R{{0.33}\linewidth}|}
    \hline
        \textbf{Metadata} & \textbf{Description} & \textbf{Codes} \\
        \hline
       Text Type  & Category or position of text element used in visualization design & Titles, Subtitles, Annotations, Captions, Axes, Legends, Paragraphs\\
       \hline
        Non-Data Visual Elements & Visual elements associated with the text, if any & Arrows, Circles, Logos, Icons, Lines, Rectangles \\
        \hline
        Color Use & Role of color in the text element, if any & Encoding, Highlight, Style \\
         \hline
    \end{tabular}
    \caption{Coding types other than text functions, which are defined in \cref{tab:functions}.
    Not all text elements contain visual elements or color.  
    }
    \label{tab:text_metadata}
\end{table}

We then coded the presence of functions in each text element.
To maintain rigor, the coding process was carried out iteratively, with a focus on achieving high reliability across all function codes \cite{hruschka2004reliability, nili2020approach}. Initially, two coders independently analyzed a random subset of 20 designs, and interrater reliability (IRR) was calculated \cite{gamer2012irr}. If the IRR did not surpass a threshold of 75\% of codes with $\kappa > 0.8$,
the coders discussed discrepancies and refined the codebook.
This iterative process continued until the threshold was reached.

After two rounds of subset coding, we surpassed the reliability threshold, achieving an average kappa value of 0.97 across the codes. More than 75\% of codes showed complete agreement between coders, although emotional engagement proved slightly more subjective, with a kappa value of 0.62. 
Each coder than coded the entire corpus independently, with discrepancies resolved through in-depth discussions among all four authors. 
The final codebook reflects these deliberations, providing a reliable framework for future analysis.

\section{Text Functions}

\begin{table*}[ht!]
\centering
\def\arraystretch{1.2}
\caption{Text functions in visualizations, constructed with a verb to describe the text's action and a noun to capture what it acts upon. ``Identify,'' ``Present,'' ``Compare,'' and ``Summarize'' are verbs borrowed from prior work \cite{rahman2024qualitative, brehmer2013multi}. Additional examples can be found in supplementary materials.
}
\begin{tabular}{|lR{0.11\linewidth}R{0.7\linewidth}|}
\hline
\textbf{Verb} &
\textbf{Noun} &
  \textbf{Definition} \\
  \hline
\raisebox{-0.2em}{\includegraphics[height=0.9em]{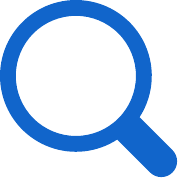}}~\textsc{Identify}  &
\raisebox{-0.2em}{\includegraphics[height=0.9em]{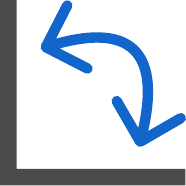}}~\textsc{Mappings}  &
    Communicates conventions for reading the charts, i.e., the relationship between data point(s) and corresponding visual elements or structures, such as position, color, shape, or other visual channels. \\
  \hline
\raisebox{-0.2em}{\includegraphics[height=0.9em]{figs/icons/identify.pdf}}~\textsc{Identify} &
\raisebox{-0.2em}{\includegraphics[height=0.9em]{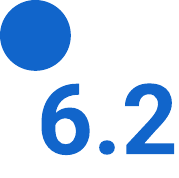}}~\textsc{Values} &
  Directly identifies and labels all data points with the relevant value, category, or other point-specific information. Serves as a reference for data values, rather than selective emphasis or comparison. \\
  \hline
\raisebox{-0.2em}{\includegraphics[height=0.9em]{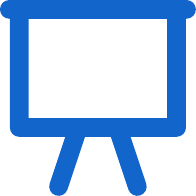}}~\textsc{Present} &
\raisebox{-0.2em}{\includegraphics[height=0.9em]{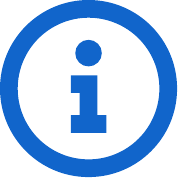}}~\textsc{Metadata} &
  Provides information about the source of the data, the transformations applied to the data, the visualization elements, its purpose, the people involved in its creation, and its intended audience. Includes definitions of a data category, variable, or other terms used. \\
  \hline
\raisebox{-0.2em}{\includegraphics[height=0.9em]{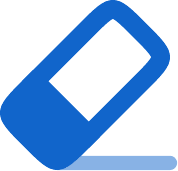}}~\textsc{Replace} &
\raisebox{-0.2em}{\includegraphics[height=0.9em]{figs/icons/mappings.pdf}}~\textsc{Mappings} &
  Provides an unconventional representation for a conventionally presented element used to establish the basic structure of data encoding when such an element (e.g., axis, legend) is omitted from the display. \\
  \hline
\raisebox{-0.2em}{\includegraphics[height=0.9em]{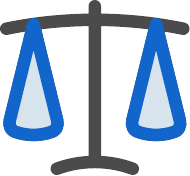}}~\textsc{Compare} &
\raisebox{-0.2em}{\includegraphics[height=0.9em]{figs/icons/mappings.pdf}}~\textsc{Mappings} &
  Translates a data mapping into more understandable or contextually relevant terms by rephrasing or interpreting technical, complex, or abstract data mappings into more relational language. \\
  \hline
\raisebox{-0.2em}{\includegraphics[height=0.9em]{figs/icons/compare.pdf}}~\textsc{Compare} &
\raisebox{-0.2em}{\includegraphics[height=0.9em]{figs/icons/values.pdf}}~\textsc{Values} &
  Describes relationships among and between data points through direct comparison of points or groups. Text may also highlight one or multiple data points in comparison to the overall dataset. \\
  \hline
\raisebox{-0.2em}{\includegraphics[height=0.9em]{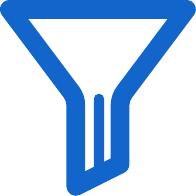}}~\textsc{Summarize} &
\raisebox{-0.2em}{\includegraphics[height=0.9em]{figs/icons/values.pdf}}~\textsc{Values} &
  Describes relationships among and between data points through aggregation (e.g., average) or mathematical function (e.g., addition). Text may also group or filter points along a given dimension. \\
  \hline
\raisebox{-0.2em}{\includegraphics[height=0.9em]{figs/icons/summarize.pdf}}~\textsc{Summarize} &
\raisebox{-0.2em}{\includegraphics[height=0.9em]{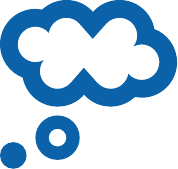}}~\textsc{Concepts} &
  Provides a high-level summary of some or all aspects of the chart; text can either provide information synthesis (subcode: \textsc{Synthesis}) or describe the variables displayed (subcode: \textsc{Variables}). Augmented by an additional taxonomy of rhetorical strategies \cite{leigh1994use, hao2024design}. \\
  \hline
\raisebox{-0.2em}{\includegraphics[height=0.9em]{figs/icons/present.pdf}}~\textsc{Present} &
\raisebox{-0.2em}{\includegraphics[height=0.9em]{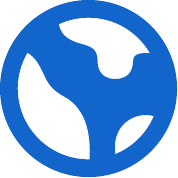}}~\textsc{Context} &
  Integrates contextual information, including background knowledge about the world (such as geographic, cultural, and political relationships), knowledge about current events, and domain-specific information stemming from expertise in a particular field of research or scholarship. \\
  \hline
\raisebox{-0.2em}{\includegraphics[height=0.9em]{figs/icons/present.pdf}}~\textsc{Present} &
\raisebox{-0.2em}{\includegraphics[height=0.9em]{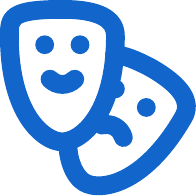}}~\textsc{Valenced Subtext} &
  Promotes an emotional response, appealing to emotions, values, and personal experience. Conveys an emotional tone that would be diminished or lost in a more neutral phrasing. \\
  \hline
\end{tabular}
 \label{tab:functions}
\end{table*}

\begin{figure*}[h!]
    \centering
    \includegraphics[width=\linewidth]{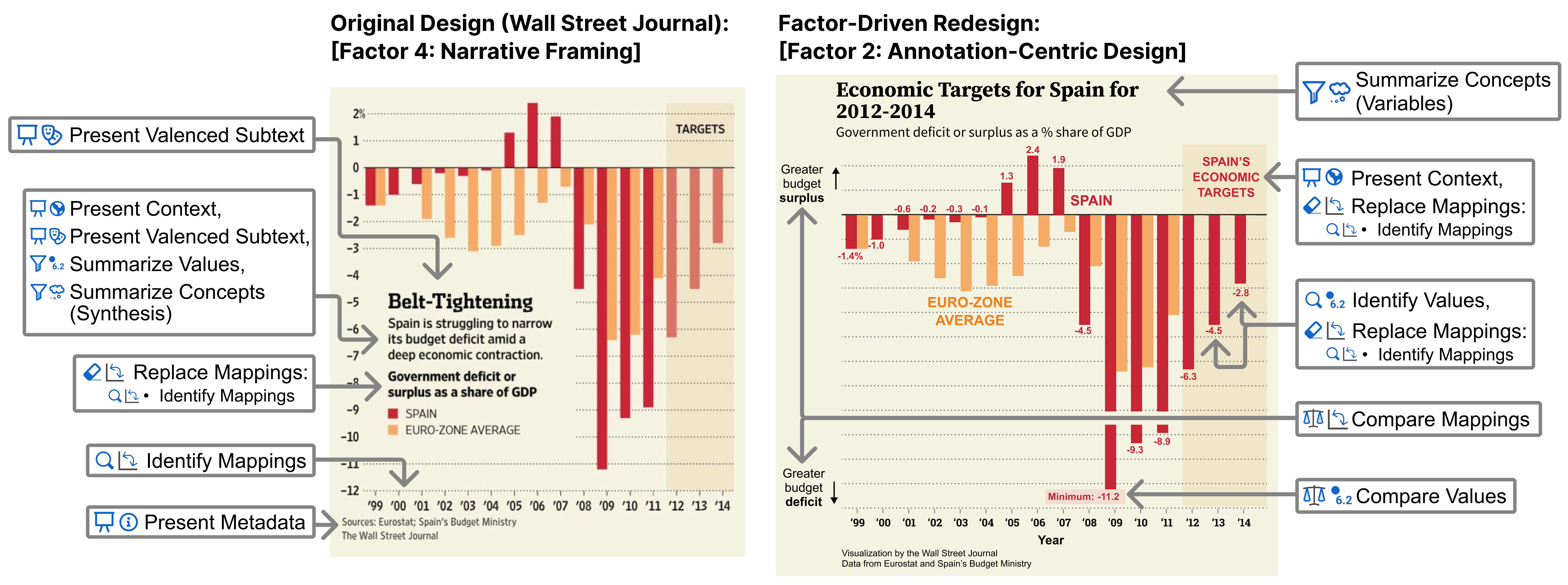}
    \caption{
     Two versions of a Wall Street Journal (WSJ) visualization demonstrate how changing text elements and the functions they perform can affect the overall information provided by the design.  \textbf{Left}: Original WSJ design~\cite{img241}. Most of the text elements contribute to the narrative framing of the data. This design falls under \fFourNarrative (see \cref{sec:factor4}). \textbf{Right}: Visualization redesigned by this paper's authors as a more annotation-centric (\fTwoAnnotation; see \cref{sec:factor2}) design. Data synthesis was also replaced with a neutral summary. This design replaces the legend and y-axis with annotations providing direct values.
     Many redundant functions are not labeled due to space considerations.
     }
    \label{fig:functions}
\end{figure*}

To clarify the relationship between text and visual elements and identify recurring design patterns across visualizations, we developed a framework comprising ten distinct text functions. 
This framework was informed by existing research on visualization tasks and text taxonomies and was further refined through open coding of a representative set of existing visualizations. 
 \cref{tab:functions} summarizes  the resulting functions, and  \cref{fig:functions} illustrates them.
Frequency information for each function can be found in \cref{tab:function_freq}; text elements could serve multiple functions.

\subsection{Text Function Framework}
\label{sec:text_functions}

Text elements that \identifyMappings communicate the conventions for interpreting the chart. \axes and \legends frequently serve this function by mapping data to visual channels such as position or hue. However, this function is not always handled by traditional components. On the right-hand side of \cref{fig:functions}, no formal y-axis exists. Instead,  
data labels (the number \annotations on the bars) take on the \identifyMappings function and show how height relates to numeric value. 

These same labels also serve a second function: \identifyValues. By providing a complete set of numeric values for each bar in the `Spain' category, they allow readers to retrieve exact values without estimating. Crucially, \identifyValues only applies when all relevant values are labeled; selective emphasis, such as tagging only an outlier, would fall under \compareValues. 

Some text provides context beyond the chart’s encodings or data.
The function \presentMetadata captures when text includes details like data sources, collection methods, or the intended audience. This function often appears in \captions, such as the note in \cref{fig:functions} that reads, ``Data from Eurostat and Spain’s Budget Ministry,'' supporting transparency and credibility.

During open coding, we identified another distinct function: \replaceMappings.
Unlike text that complements conventional encodings, this function substitutes for them entirely. 
In the right of \cref{fig:functions}, for example, the y-axis is omitted. Instead, text on each bar communicates the encoded values, replacing what the axis would have shown. 
Similarly, \annotations such as `Spain' and `Euro-Zone Average' take the place of \legends. Even on the left side of \cref{fig:functions}, the legend title ``Government deficit or surplus as a share of GDP'' doubles as a replacement for a y-axis label. These examples illustrate how text can step in when conventional elements are minimized or removed.

Other text elements work to interpret or rephrase chart encodings for the viewer. 
The \compareMappings function translates visual encodings into more conceptual or accessible terms.
For instance, the phrase, ``Greater budget surplus,'' on the right-hand side of \cref{fig:functions} offers a qualitative interpretation of the vertical position. 
Unlike \identifyMappings, which describe standard mappings,
\compareMappings represents the \textit{meaning} of the encoding in the context of the data.

When text highlights relationships between specific data points, it performs the \compareValues function. 
This can be as part of a longer phrase or as a single data label. 
This includes callouts to outliers, extremes, or other noteworthy comparisons. 
For instance, an \annotationType identifying the minimum value (as on the right of \cref{fig:functions}) emphasizes how it stands apart from the rest, even if other points are left unlabeled.

Text can also summarize broader patterns. The function \summarizeValues applies when text aggregates or calculates across multiple points, such as when describing an average or trend. 
On the left of \cref{fig:functions}, the \subtitle, ``Spain is struggling to narrow its budget deficit amid a deep economic contraction,'' reflects the full set of data for Spain and conveys a high-level trend.
Both \summarizeValues and \compareValues have a set of more precise subfunctions drawn from  a low-level task taxonomy \cite{amar2005low}.

The same \subtitle also serves another function: \summarizeConcepts.
This function captures broader conceptual framing or synthesis, often seen in \titles or \subtitles. 
Through close review of title-specific studies and taxonomies \cite{liu2023autotitle, kong2018frames, kong2019trust}, we refined this category and identified two subtypes.
\summarizeConceptsSyn involves interpretation or synthesis, distilling the chart into a takeaway (e.g., left of \cref{fig:functions}: ``Spain is struggling...''). 
\summarizeConceptsVar, by contrast, lists the variables or chart contents \textit{without} added interpretation (e.g., right of \cref{fig:functions}: ``Economic Targets...''). 
We also examined rhetorical strategies used in \summarizeConcepts, such as puns, associations, exaggerations, repetitions, etc. \cite{leigh1994use, hao2024design}.

Finally, text can introduce information not visible in the chart. 
\presentContext includes external background, such as domain knowledge or current events.
In the left-hand chart in \cref{fig:functions}, the phrase ``deep economic contraction'' situates the data in a broader economic narrative. Even brief text, like the ``Targets'' \annotationType, can signal contextual knowledge that these are not just values, but policy goals.

The \presentValence function promotes emotionally charged or value-laden language. 
For instance, describing Spain as ``struggling'' introduces an affective tone not inherent in the data presentation otherwise. 
To inform our definition and use of this function, we reviewed studies on affect in visualization design \cite{lee2022affective, lan2023affective}. 
Since we did not have access to designer intents, we focused instead on the text’s \textit{potential} to elicit an emotional or affective response in viewers. 
A useful heuristic is whether the text could be reasonably rewritten in a more neutral form; if so, the original likely carries affective weight.

\begin{table*}
\centering
\def\arraystretch{1.2}
\caption{
Percent of designs ($n = 120$) containing each function, faceted by text type. Designs could contain multiple types of text serving the same function. Cells are shaded based on values \textit{within} each column, with overall occurrence percentages highlighted in blue and type-specific percentages in gray. Annotations and subtitles were the most functionally diverse types of text. 
}
\begin{tabular}{|l|r|r|r|r|r|r|r|r|}
\hline
\textbf{Function} &
  \multicolumn{1}{R{0.065\linewidth}|}{\textbf{Percent of Corpus}} &
  \multicolumn{1}{l|}{\textbf{\titles}} &
  \multicolumn{1}{l|}{\textbf{\subtitles}} &
  \multicolumn{1}{l|}{\textbf{\annotations}} &
  \multicolumn{1}{l|}{\textbf{\captions}} &
  \multicolumn{1}{l|}{\textbf{\axes}} &
  \multicolumn{1}{l|}{\textbf{\legends}} &
  \multicolumn{1}{l|}{\textbf{\paragraphs}} \\
\hline
\raisebox{-0.2em}{\includegraphics[height=0.9em]{figs/icons/identify.pdf}}~\raisebox{-0.2em}{\includegraphics[height=0.9em]{figs/icons/mappings.pdf}}~Identify Mappings &
  \cellcolor[HTML]{9FC5E8}100\tabpercent &
  \cellcolor[HTML]{F8F8F8}11\tabpercent &
  \cellcolor[HTML]{F1F1F1}9\tabpercent &
  \cellcolor[HTML]{CCCCCC}58\tabpercent &
  \cellcolor[HTML]{FDFDFD}3\tabpercent &
  \cellcolor[HTML]{CCCCCC}75\tabpercent &
  \cellcolor[HTML]{CCCCCC}44\tabpercent &
  \cellcolor[HTML]{F8F8F8}2\tabpercent{} \\
\hline
\raisebox{-0.2em}{\includegraphics[height=0.9em]{figs/icons/identify.pdf}}~\raisebox{-0.2em}{\includegraphics[height=0.9em]{figs/icons/values.pdf}}~Identify Values &
  \cellcolor[HTML]{D4E5F5}56\tabpercent &
  \cellcolor[HTML]{FFFFFF}0\tabpercent &
  \cellcolor[HTML]{FEFEFE}2\tabpercent &
  \cellcolor[HTML]{D2D2D2}52\tabpercent &
  \cellcolor[HTML]{FFFFFF}0\tabpercent &
  \cellcolor[HTML]{FDFDFD}4\tabpercent &
  \cellcolor[HTML]{FEFEFE}1\tabpercent &
  \cellcolor[HTML]{FFFFFF}1\tabpercent{} \\
\hline
\raisebox{-0.2em}{\includegraphics[height=0.9em]{figs/icons/present.pdf}}~\raisebox{-0.2em}{\includegraphics[height=0.9em]{figs/icons/metadata.pdf}}~Present Metadata &
  \cellcolor[HTML]{C5DCF2}68\tabpercent &
  \cellcolor[HTML]{FCFCFC}6\tabpercent &
  \cellcolor[HTML]{ECECEC}12\tabpercent &
  \cellcolor[HTML]{FBFBFB}7\tabpercent &
  \cellcolor[HTML]{CCCCCC}53\tabpercent &
  \cellcolor[HTML]{FFFFFF}0\tabpercent &
  \cellcolor[HTML]{FBFBFB}4\tabpercent &
  \cellcolor[HTML]{F8F8F8}2\tabpercent{} \\
\hline
\raisebox{-0.2em}{\includegraphics[height=0.9em]{figs/icons/replace.pdf}}~\raisebox{-0.2em}{\includegraphics[height=0.9em]{figs/icons/mappings.pdf}}~Replace Mappings &
  \cellcolor[HTML]{D4E5F5}56\tabpercent &
  \cellcolor[HTML]{FAFAFA}8\tabpercent &
  \cellcolor[HTML]{F3F3F3}8\tabpercent &
  \cellcolor[HTML]{DFDFDF}38\tabpercent &
  \cellcolor[HTML]{FFFFFF}1\tabpercent &
  \cellcolor[HTML]{FDFDFD}4\tabpercent &
  \cellcolor[HTML]{FDFDFD}2\tabpercent &
  \cellcolor[HTML]{F8F8F8}2\tabpercent{} \\
\hline
\raisebox{-0.2em}{\includegraphics[height=0.9em]{figs/icons/compare.pdf}}~\raisebox{-0.2em}{\includegraphics[height=0.9em]{figs/icons/mappings.pdf}}~Compare Mappings &
  \cellcolor[HTML]{FFFFFF}19\tabpercent &
  \cellcolor[HTML]{FFFFFF}0\tabpercent &
  \cellcolor[HTML]{FFFFFF}1\tabpercent &
  \cellcolor[HTML]{F6F6F6}12\tabpercent &
  \cellcolor[HTML]{FFFFFF}0\tabpercent &
  \cellcolor[HTML]{FEFEFE}2\tabpercent &
  \cellcolor[HTML]{FDFDFD}2\tabpercent &
  \cellcolor[HTML]{F8F8F8}2\tabpercent{} \\
\hline
\raisebox{-0.2em}{\includegraphics[height=0.9em]{figs/icons/compare.pdf}}~\raisebox{-0.2em}{\includegraphics[height=0.9em]{figs/icons/values.pdf}}~Compare Values &
  \cellcolor[HTML]{EEF5FB}34\tabpercent &
  \cellcolor[HTML]{FCFCFC}5\tabpercent &
  \cellcolor[HTML]{F7F7F7}6\tabpercent &
  \cellcolor[HTML]{EBEBEB}24\tabpercent &
  \cellcolor[HTML]{FFFFFF}1\tabpercent &
  \cellcolor[HTML]{FFFFFF}1\tabpercent &
  \cellcolor[HTML]{FEFEFE}1\tabpercent &
  \cellcolor[HTML]{F1F1F1}3\tabpercent{} \\
\hline
\raisebox{-0.2em}{\includegraphics[height=0.9em]{figs/icons/summarize.pdf}}~\raisebox{-0.2em}{\includegraphics[height=0.9em]{figs/icons/values.pdf}}~Summarize Values &
  \cellcolor[HTML]{EDF4FB}35\tabpercent &
  \cellcolor[HTML]{FAFAFA}8\tabpercent &
  \cellcolor[HTML]{ECECEC}12\tabpercent &
  \cellcolor[HTML]{F5F5F5}13\tabpercent &
  \cellcolor[HTML]{FFFFFF}0\tabpercent &
  \cellcolor[HTML]{FFFFFF}0\tabpercent &
  \cellcolor[HTML]{FEFEFE}1\tabpercent &
  \cellcolor[HTML]{EAEAEA}4\tabpercent{} \\
\hline
\raisebox{-0.2em}{\includegraphics[height=0.9em]{figs/icons/summarize.pdf}}~\raisebox{-0.2em}{\includegraphics[height=0.9em]{figs/icons/concepts.pdf}}~Summarize Concepts &
  \cellcolor[HTML]{AECEEC}88\tabpercent &
  \cellcolor[HTML]{CCCCCC}78\tabpercent &
  \cellcolor[HTML]{CCCCCC}30\tabpercent &
  \cellcolor[HTML]{F5F5F5}13\tabpercent &
  \cellcolor[HTML]{F9F9F9}7\tabpercent &
  \cellcolor[HTML]{FFFFFF}1\tabpercent &
  \cellcolor[HTML]{FDFDFD}2\tabpercent &
  \cellcolor[HTML]{F1F1F1}3\tabpercent{} \\
\hline
\raisebox{-0.2em}{\includegraphics[height=0.9em]{figs/icons/present.pdf}}~\raisebox{-0.2em}{\includegraphics[height=0.9em]{figs/icons/context.pdf}}~Present Context &
  \cellcolor[HTML]{F0F6FC}32\tabpercent &
  \cellcolor[HTML]{FEFEFE}3\tabpercent &
  \cellcolor[HTML]{ECECEC}12\tabpercent &
  \cellcolor[HTML]{F1F1F1}18\tabpercent &
  \cellcolor[HTML]{FBFBFB}5\tabpercent &
  \cellcolor[HTML]{FEFEFE}2\tabpercent &
  \cellcolor[HTML]{FDFDFD}2\tabpercent &
  \cellcolor[HTML]{CCCCCC}8\tabpercent{} \\
\hline
\raisebox{-0.2em}{\includegraphics[height=0.9em]{figs/icons/present.pdf}}~\raisebox{-0.2em}{\includegraphics[height=0.9em]{figs/icons/valence.pdf}}~Present Valenced Subtext &
  \cellcolor[HTML]{F8FBFE}25\tabpercent &
  \cellcolor[HTML]{F2F2F2}21\tabpercent &
  \cellcolor[HTML]{F3F3F3}8\tabpercent &
  \cellcolor[HTML]{FFFFFF}2\tabpercent &
  \cellcolor[HTML]{FFFFFF}0\tabpercent &
  \cellcolor[HTML]{FFFFFF}0\tabpercent &
  \cellcolor[HTML]{FFFFFF}0\tabpercent &
  \cellcolor[HTML]{F8F8F8}2\tabpercent{} \\
\hline
\end{tabular}
  \label{tab:function_freq}
\end{table*}

\subsection{Comparison to Other Frameworks}

In comparison to existing frameworks, our approach aligns with prior taxonomies while also offering meaningful extensions. 
Several of our functions, such as \identifyMappings, \summarizeValues, and \compareValues, align with the action-oriented verbs in Rahman et al.'s annotation taxonomy \cite{rahman2024qualitative}.
Similarly, Lundgard and Satyanarayan’s four-level semantic model \cite{lundgard2021accessible} captures broader categories like contextual or encoded information, which correspond to our \presentContext and \identifyMappings functions. 
Our framework builds on these foundations by introducing finer-grained, context-sensitive functions that interpret text in relation to its role within the overall chart design. 
This added specificity enables more detailed analysis of design choices and more precisely reflects how text operates as an integrated part of visualization structure, rather than as isolated content.

Our schema also diverges from other taxonomies in its inclusion of functions such as \replaceMappings and \presentValence, which capture nuanced roles of text that are largely absent from existing taxonomies. These functions were developed through the open coding procedures; existing taxonomies were unable to capture the role of the respective text.
For example, while the semantic levels address text that describes visual encodings, they do not account for cases where text replaces traditional components, such as axes or legends, as captured by \replaceMappings. 
As another example, \compareMappings bridges L1 (elemental and encoded) and L2 (statistical and relational) levels \cite{lundgard2021accessible} by connecting abstract encodings to relational terms. 
The potential for text to evoke emotion, represented by \presentValence,  goes beyond the scope of most taxonomies.
Additionally, our framework captures the fact that even seemingly simple text elements can perform \textit{multiple} functions, something not commonly labeled in other schemes.

\subsection{Exploratory Analysis Findings}
\label{sec:other_findings}

We analyzed the prevalence and representation of different functions; further details can be found in supplementary materials.
This analysis addresses our first research question: \textbf{What are the \textit{functions} of text in visualization designs?} 
We examine how function frequencies and usage patterns align with established conventions in text design, and to what extent the function framework reveals additional or previously undercharacterized aspects of how text is used in visualizations.

\subsubsection{Characterizing Titles and Subtitles}
\label{sec:titles}

Prior research indicates that titles provide a strong initial impression of the visualization’s content and meaning \cite{borkin2015beyond, kong2018frames, kong2019trust}, but the majority of prior work does not provide detailed analysis of the construction or interplay of titles and subtitles. Consider the two examples in \cref{fig:functions}: 

\begin{tightItemize}
    \item[T1:] ``Economic Targets for Spain for 2012-2014: Government deficit or surplus as a \% share of GDP''
    \item[T2:] ``Belt-Tightening: Spain is struggling to narrow its budget deficit amid a deep economic contraction.''
\end{tightItemize}

\noindent
These examples exhibit text function structure that is common within our collection.
Type T1 is labeled in our framework as \summarizeConceptsVar and we find that in almost every case (94\%), this type contains noun phrases but no verbs.
In contrast, type T2 is labeled as \summarizeConceptsSyn, and these often contain verbs (68\%).   

T2 demonstrates another pattern common to some news visualizations: a short title exhibiting a rhetorical move, in this case a metaphor (``Belt-tightening''), as a bid to capture the reader's attention.
Using a taxonomy of headline strategies from advertising \cite{leigh1994use} and data-driven journalism \cite{hao2024design}, we analyzed rhetorical devices such as puns, associations, exaggerations, word omissions, and repetition, alongside broader distinctions between formal and informal registers \cite{iso_registers}.
We observed much higher rates of puns and metaphors in short \titles (three or fewer words; 62\%) compared to \subtitles or longer \titles (3\%). For example, the \titleType ``Falling Rains'' (Image 243) encapsulates both a variable (rain) and a data trend (declining rainfall), leveraging wordplay in the form of a pun to reinforce meaning and engage the reader. 
In contrast, \subtitles rarely used distinct headline rhetorical strategies, suggesting that \subtitles serve a different role, often providing additional context rather than relying on wordplay or rhetorical techniques.

\titles that used rhetorical strategies often also \presentValence through associations or wordplay, which is a part of the ``Narrative'' factor. These short \titles are typically paired with longer \subtitles.
Additionally, by using the double meaning in a word, \titles can simultaneously perform \summarizeConceptsSyn and \summarizeConceptsVar.

Although \titles are a common element in visualizations, 26 images in our corpus omitted them entirely. In these cases, text information was primarily conveyed through \annotations, with an average of 3.65 \annotations per image (\textit{SD} = 2.56). 
These \annotations were most commonly used to \identifyValues, though some engaged in data synthesis, summarization, or comparison. 
\annotations were the predominant location for \compareValues, allowing designers to position comparative statements directly alongside relevant data. This pattern reflects an overlap between the \annotationFactor and \narrative
factors described in \cref{sec:factors}. \annotations can both serve as substitutes for conventional chart elements and contribute to storytelling by emphasizing comparisons.

\subsubsection{Replacing Axes and Legends}
\label{sec:replace_mappings}

Across our corpus, 56\% of designs omitted at least one traditional mapping element, instead using \replaceMappings.
This function was most common in \annotations but also appeared in \titles and \subtitles, where text elements assumed roles typically associated with axis labels. 
\axes were replaced slightly more often than \legends, likely influenced by their more frequent presence in visualizations; 31 images replaced \legend, and 41 images replaced \axes. Six designs replaced both a \legend and an \axis with other text elements.
We observed three primary representations of \replaceMappings (see \cref{fig:teaser}).

The most common use of \replaceMappings replaced \legends with \annotations ($n = 28$), sometimes incorporating the use of color directly into the text ($n = 14$).
For example, the right image in \cref{fig:teaser} uses color in the ``Democrat'' and ``Republican'' \annotations, allowing the designer to avoid finding suitable space for a \legend.
This method is a form of direct labeling \cite{ajani2021declutter, knaflic2015storytelling, franconeri2021science}, reducing the need for users to reference a \legend. 
In cases where color was not used, the data points were directly labeled, or the color encoding was simply described in the text, as in the case of Image 403 in \cref{fig:factor_loadings}. An \annotationType explicitly states that red shading indicates uncertainty, eliminating the need for a separate legend box. 
This method may be particularly useful when space is limited, or the color encodings require more explanation than a \legend affords. 

\titles or \subtitles could also be used to present \axis information ($n = 17$), eliminating the need for a separate axis label. 
This approach could reduce redundant text while maintaining a neutral framing of the data.  
For example, the center image in \cref{fig:teaser} uses the \subtitle ``Height in metres, \textit{(year of completion)},'' which provides the same information as an axis label but in a more prominent location. 
This method aligns with minimalist design practices, though it requires viewers to synthesize information across different text elements rather than relying on a localized \axis label. 

\axes could also be replaced with direct data labels in the form of \annotations ($n = 14$).
Instead of requiring viewers to estimate values along a scale, designs used numeric values positioned directly within or beside data points.
This strategy can be seen in the left image of \cref{fig:teaser} where the percent values are directly labeled and the 0 to 100 y-axis is only implied. In this case, the design prioritizes precision and reduces the need for external reference points. 
However, it can also increase text density, which may become overwhelming in data-dense visualizations. For example, in \cref{fig:functions}, the decision to label Spain’s data directly required widening the design.

\subsubsection{Single Text,  Multiple Functions}
\label{sec:multifunctional}

Within our corpus, 39\% of text components served multiple functions, with \annotations being the most frequently multifunctional (66\%), followed by \subtitles (58\%) and \titles (51\%). 
These text elements often worked to construct a visualization’s narrative structure rather than just supporting visual encodings.

Several common function pairings emerged from our analysis. 
\summarizeValues and \compareValues frequently co-occurred, with 58\% of \summarizeValues instances also involving \compareValues. The combination of these functions suggests that data synthesis often involved both aggregation and direct comparisons. This kind of combination may allow for complex data operations and calculations to be summarized in a single sentence or phrase.

Similarly, \presentValence almost always (90\%) appeared in combination with \summarizeConcepts and, to a lesser extent (28\%), \presentContext. This pattern highlights the role of some \titles and \subtitles in shaping emotional framing and narrative context rather than presenting data in a purely neutral manner, as we explored in \cref{sec:titles}. The frequency of this function combination highlights the strategies used by designs to engage with readers.

Text that expressed the \compareMappings functions also frequently expressed  \summarizeValues (22\%), It was also common for \compareMappings to co-occur (34\%) with \replaceMappings;  in these cases, the design would typically omit one or both \axes in favor of \annotations.

Some functions tended to operate independently of others. \presentMetadata was largely independent of other text functions, with its highest co-occurrence being 11\% with \identifyMappings, and even lower with all other functions.
These findings suggest that although some text functions may combine to form layered interpretative structures, others remain more specialized or independent, serving distinct roles within the visualization.

\section{Design Patterns Based on Text Functions}
\label{sec:factors}

To investigate how text functions and design features interact to form latent constructs — 
underlying patterns that emerge from correlated text and design elements in visualizations — 
we conducted an exploratory factor analysis (EFA) \cite{gorsuch1988exploratory, thompson2004exploratory} with varimax rotation in R Studio using the \texttt{psych} package 
\cite{revelle2024psych, Rstudio}. 
This analysis addresses our second research question: \textbf{What text design patterns emerge across visualizations, and which functions are engaged in each pattern?}
Similar clustering analyses have been performed in prior work
\cite{sarikaya2019dashboards, srinivasan2024zoo, bach2023dashboard}.
 
We included binary features based on: our proposed text \textbf{functions}, text \textbf{type} (e.g., \titleType), image \textbf{domain} (e.g., News), \textbf{color} use (e.g., encoding), \textbf{visual elements} associated with the text (e.g., arrows), and normalized \textbf{word count} measures (e.g,. total word count). See \cref{tab:text_metadata} for more information on the text metadata categories.
\identifyMappings was not included in this analysis since it appeared in every design.
Further details can be found in the supplementary materials. 

\subsection{Factor Selection}

We examined the feature correlation matrix to identify relationships between text properties. As expected, there were strong correlations among certain variables; \presentMetadata and \captions had the highest correlation (\textit{r} = 0.74).
At the same time, many features were only weakly correlated, reinforcing the need for a multifactor solution.
To determine the number of factors, we used Bayesian information criterion (BIC), model complexity, and eigenvalues~\cite{hayton2004factor}. These measures collectively supported a 4-factor model, which best captured variation in text use  while preserving factor independence and interpretability. 

\subsection{Factor Analysis Evaluation}
To assess the reliability of the factor analysis in identifying latent textual constructs within our corpus, we deployed several validation approaches, including blind expert categorization, factor reliability assessment through resampling, and permutation testing.

As an initial evaluation, three coauthors -- blinded to the factor analysis results -- independently categorized a subset of visualizations from the corpus. Using the four identified factors as a reference, they predicted which factor each visualization would be most associated with. 
This blind categorization was performed for five images, and their classifications showed strong alignment with the factor loadings derived from the analysis, providing initial support for factor validity.

We further evaluated factor reliability through resampling and permutation testing. 
For resampling, we randomly selected 75\% of images in each iteration and re-ran the 4-factor analysis, extracting factor loadings. The 75\% subset size balanced sample representativeness and computational feasibility. We repeated this process for 50 iterations, determined empirically through stabilization of intraclass correlation (ICC) values. ICC values showed minimal change beyond 50 iterations, indicating convergence. After 50 iterations, ICC values for all four factors exceeded 0.93, confirming high factor reliability.

Additionally, we tested robustness by removing the top 10\% of images based on their factor loadings to check if a small subset of highly weighted images disproportionately influenced the factor structure. The resulting factors remained comparable to those from the full dataset, supporting the validity of our extracted factors.
These findings reinforce that the identified factors reflect underlying text relationships in visualizations rather than being artifacts of specific samples or outliers. 

Together, these validation steps ensure that the factor analysis reliably captures meaningful text-based constructs within the visualization corpus, reinforcing the robustness of our findings.

Our exploratory factor analysis identified four distinct approaches to text design;
examples of each factor along with its loadings can be seen in \cref{fig:factor_loadings}. 
We assigned names to these factors to make them easier to reference and understand \cite{sarikaya2019dashboards}, while acknowledging that the names may not capture every specific variable perfectly.
Variables with loadings above 0.6 were considered strongly associated; variables with loadings below 0.3 were considered weakly associated.
We supplemented the factor analysis with additional exploratory analyses to identify cross-factor findings and practices.

The right-hand version of the visualization shown in \cref{fig:functions} illustrates the use of these factors. The original version on the left aligns with \narrative. We converted this design to align with the variables that load highly on \annotationFactor,  by shifting the emphasis toward \identifyValues and \annotationType.

\begin{figure*}
    \centering
    \includegraphics[width=\linewidth]{figs/factors.pdf}
    \caption{Four factors identified in exploratory factor analysis, including the relevant items, their loadings, and example images for each factor. 
    These factors represent four patterns of text design practices in visualizations: \attribution, \annotationFactor, \visual, and \narrative.
    Loading scores indicate the strength of each variable’s association with the factors.
    Example images shown are also available in supplementary materials for reference.
}
    \label{fig:factor_loadings}
\end{figure*}

\subsection{Factor 1: Attribution and Variables}
\label{sec:factor1}

We refer to \fOne as \attribution, since this factor has relatively strong loadings from \presentMetadata and \summarizeConceptsVar.
Overall, loadings indicate that text primarily plays a neutral, factual role. 

The highest loading item in this factor is \captions, which overwhelmingly served to \presentMetadata (93\%), following standard journalistic conventions for citing sources and providing attribution \cite{hao2024design}. \titles were moderately associated with this factor, most of performed \summarizeConceptsVar, providing no additional framing to the data (87\%). \subtitles also had a moderate loading on this factor. Overall, \attribution appears to capture common text design practices in journalism visualizations, with a moderate loading from news sources.
Conversely, blog and media-based visualizations were moderately negatively associated with \fOne.

One common design pattern in this factor is highly neutral text, where text functions almost entirely \presentMetadata and \summarizeConceptsVar.
For example, in Image 55 in \cref{fig:factor_loadings}, nearly all non-axis text is dedicated to these neutral functions, resulting in a design where text serves as an organizational and attributional tool rather than contributing to interpretation.

A second pattern within this factor
uses long \subtitles to provide additional context; \subtitle word length also had a moderate loading on \fOne. 
For instance, in Image 239 in \cref{fig:factor_loadings}, the \titleType is the elliptical `Buried Treasure,' and the \subtitle states, ``Afghanistan has numerous mineral deposits that it is trying, against considerable challenges, to develop.'' The \titleType offers only minimal insight into the data, but the \subtitle elaborates on the content displayed.

Note that this \subtitle introduces contextual information; most examples within this factor contain text that is largely but not 100\% neutral.
These sets of loadings suggest that \fOne captures a broader ``fact-based'' focus to charts. Designs in the \attribution factor tended to have an emphasis on neutral summaries and statements of attribution but could still incorporate additional context or data synthesis through the use of longer \subtitles.

\subsection{Factor 2: Annotation-Centric Design}
\label{sec:factor2}

We refer to \fTwo as \annotationFactor, since it encompasses a design approach characterized by a high density of \annotations, the use of text to \identifyValues, and substitutions in the form of \replaceMappings. 
Designs in this factor tended to embed details directly within the chart itself rather than using longer \titles to describe data features.  

This pattern reveals a possible design strategy that prioritizes in situ explanation, with positive loadings for both \annotations (0.65) and \annotationType word count (0.65), alongside a negative loading for \titleType word count (-0.63). 
By distributing a possibly lengthy title within the chart, the text density of the design may not be greatly increased by added \annotations. 
Based on the moderate loading of \identifyValues (0.53) also within \fTwo, these annotations likely also provide additional data precision in many cases. 

\replaceMappings had a moderate loading (0.33) in this factor (see \ref{sec:replace_mappings} for more detail on this function), likely corresponding in part with the use of \identifyValues and the weak negative loading of \legends.
Government-produced visualizations also had a negative loading, which suggests that government designs may rely more on conventional data presentation methods, with fewer \annotations replacing traditional mapping elements.

Although not a major component of \fTwo, \compareMappings had a weakly positive association with the factor. \annotations tended to be the primary location for the \compareMappings function (\cref{tab:function_freq}). 
\compareMappings had a relatively equal but negative loading on \attribution, indicating that \fOne designs tend to rely on more conventional visual mappings rather than creative uses of text, such as \compareMappings or \replaceMappings.

\subsection{Factor 3: Visual Embellishments}
\label{sec:factor3}

\fThree is referred to as \visual, since it primarily captures how text interacts with stylistic elements in a design, highlighting the use of color, icons, and other graphical elements to add emphasis.
Unlike other factors, which have strong loadings from text functions, variables in this factor primarily reflect aesthetic decisions.
In addition to using color to style the text, designs in this factor incorporate embellishments, often connecting the text to a specific data point (with a circle or line) or to a higher-level concept (using logos or icons).
 
The strongest-loading item in this factor is the use of color in text for style purposes. This includes cases where text is colored for reasons beyond standard data encoding, such as matching a designer’s branding. 
For example, in Image 19 in \cref{fig:factor_loadings}, part of the \captionType is rendered in green to align with the logo of the designer group (Visual Capitalist), reinforcing the visual identity of the source. 

Word count also strongly loaded onto this factor, indicating that these designs tended to feature longer text components. On average, designs in \fThree contained 1.5 more text elements per visualization ($Mean = 9.4, SD = 3.1$) than those in other factors ($Mean = 7.9, SD = 3.6$). 
This may help explain the higher word counts observed in these designs since more elements contribute to the overall textual density. 

Beyond color and word count, the presence of additional visual elements (e.g., logos, icons, circles, and lines) was another key feature of \visual. Most visual elements (87\%) appeared alongside \annotations, with only four appearing near \captions. Visual embellishments were not used alongside \titles or \subtitles in our corpus, reinforcing the idea that these stylistic choices may be primarily applied to supporting text elements, such as annotations, rather than typically high-level text elements, such as titles.

Although text functions are essential for understanding how text supports visualization design, the combination of variables in \visual highlights that visual styling can serve a distinct role.
The presence of \fThree underscores the need for future research into how visual styling may interact with or operate separately from text function in visualization design.

\subsection{Factor 4: Narrative Framing}
\label{sec:factor4}

We refer to \fFour as \narrative, due to text usage patterns that align with data storytelling practices \cite{ren2017annotation, segel2010narrative}. 
\summarizeConceptsSyn and \presentValence had the strongest loadings ($>0.7$), followed by \summarizeValues and \compareValues ($>0.5$). \presentContext also had a moderate loading on this factor (0.34).
Visualizations associated with this factor could emphasize specific takeaways by integrating text that framed the data in relation to broader events or real-world implications. 

Variables in this factor represent a narrative approach to visualization design. The strong association with \summarizeConceptsSyn indicates these visualizations frequently contain text that distills complex information into concise summaries, and the nearly equal loading for \presentValence suggests these texts often incorporate emotional language, framing, or use of associations/puns.

The moderately strong loadings for \summarizeValues and \compareValues further characterize this factor through text elements that highlight specific data points and their relationships, sometimes including \presentContext, which had a moderately weak loading on \fFour.
Some of the association between these factors at the image level may have its roots in the specific text elements themselves; several function combinations that were characteristic of this factor also frequently combined within individual text elements as well (see \cref{sec:multifunctional}). 

Social media visualizations had a moderate negative association with this factor, suggesting that charts shared on social platforms may rely less on in-chart data synthesis. 
Although our corpus did not have a high degree of visualizations that were collected from social media, this suggests an interesting future direction of research to better understand the prevalence of theses factors across contexts.

\section{Discussion}

We addressed two major research questions: \textbf{What are the functions of text in visualization designs and which functions are engaged with specific emergent design patterns?}
By developing and applying a functional framework grounded in real-world examples, we aim to establish a baseline for how textual elements operate in practice: what they do, where they appear, and how they co-occur.
This work augments and improves on previous taxonomies by providing more specific characterization of functions, examining multiple types of text, considering interactions between text and visual elements, and capturing multifunctionality within a single text element. 

One striking result is the prevalence of multifunctional text elements. 
These elements challenge assumptions that each component of a visualization serves a single, fixed role, opening up new design questions: when is multifunctionality effective, and when might it obscure meaning? 
Our framework offers one way to answer that by supporting structured audits of existing designs through a combination of functional analyses and reader assessments or feedback. 
For example, a designer or evaluator might identify areas where additional text could clarify ambiguous encodings, or where overloaded annotations might be better split into separate components with distinct functions.

Beyond its research contributions, this framework also has the potential to support visualization designers who seek greater structure and intentionality in how they use text. 
Having a clear set of functional roles offers a way to systematically plan, revise, and evaluate text components. Designers can more effectively align their text with specific communicative goals or contexts. 

Within this paper, we demonstrate the use of factor-motivated redesigns (\cref{fig:functions}). The factors offer a structured way to modify text elements based on communicative goals. 
For example, visualizations associated with \visual may be well suited for public-facing reports and marketing materials, where color and graphical elements can enhance engagement. 
In contrast, when presenting the same data to stakeholders, visualizations that score higher on \narrative may be more appropriate, in which designers emphasize key takeaways and persuasive messaging.
Designers could intentionally shift between text design strategies depending on audience, purpose, and available space.

These patterns could also serve as starting points or templates.
By providing a structured framework of standard text functions and their co-occurrence patterns, our framework offers a foundation for assessing data visualization designs
to be sure the text is meaningful and contextually appropriate; our set of codes provides a start to a corpus that could also be used to evaluate visualizations in the future.

This work contributes to the foundation for a deeper theoretical understanding of text in visualization, moving beyond broad categorizations of text content to a functional approach that accounts for the diverse roles text plays in structuring and shaping visual information. 
The ability to differentiate designs based solely on text design practices further underscores its role as an active design element. 
By systematically defining and analyzing text functions, we contribute to a growing body of research that situates text as a fundamental design element rather than a secondary consideration. Future studies can build on and adapt this framework to continue expanding our understanding and recommendations for text design in visualizations. 

\section{Limitations and Future Work}
\label{sec:fw}

This study provides a broad analysis of text functions in visualization design, but there are several limitations to consider. 
One constraint of this work is that our corpus is primarily composed of visualizations drawn from prior research, meaning it may not fully reflect the most recent trends in text design. 
Additionally, the corpus is largely made up of news visualizations, which may limit the applicability of these findings to other contexts. 
Expanding future analyses to include a broader range of domains could provide additional insights into how text functions vary depending on audience, purpose, and platform.

Building on these findings, several directions for future research could further refine our understanding of text in visualization design. 
A targeted, domain-specific analysis of text function would provide valuable insights into how text is used in specific contexts, such as social media posts or scientific reports. For example, visualizations shared on social media may rely more heavily on captions and post text to provide context rather than integrating text directly into the chart itself. 
Interactive and multi-visualization designs (i.e., dashboards) also
warrant further investigation. A deeper understanding of text in visualization design also has the potential to provide more useful explanations of data for audiences with lower graphical literacy.

This work characterizes the current use of text in visualizations but does not assess the \textit{usability} of these choices to further validate the framework. Future work should also consider user studies to evaluate the impact of different text design choices on reader comprehension, trust, and engagement. 
Specifically, examining the role of text function on audience interpretation would help to shape guidelines or educational materials around text design for visualizations. Although this work explores the practices in existing designs, there is an additional need for researchers to establish empirical recommendations.

Another area of future work could use automation for text classification and generation. 
Current AI systems frequently omit or make errors in essential text elements such as titles, creating notable gaps in automated chart design \cite{chen2024viseval, tian2024chartgpt}.
Training an LLM on these functions and datasets could enable more efficient extraction and classification of text in visualizations. This would provide a faster way to track evolving text design practices in public charts. Additionally, AI-assisted visualization systems could also provide suggestions or guidance to designers, helping them make more informed text design choices.

\section{Conclusion}

This work provides a detailed framework for understanding text functions in visualization, expanding beyond broad classifications.
Through an analysis of 120 designs and 804 text elements, we identify ten distinct functions.
Our factor analysis further uncovers four key factors of text design, demonstrating how designers combine different textual features to support visual communication. 
By combining fine-grained text functions with broader design factors, this framework highlights practices and opportunities for intentional text integration in visualizations.
Future work should further validate this framework and explore how text functions influence reader interpretations, how function-based frameworks might inform AI-assisted design tools, and how visualization designers can use this framework to support more intentional text design.
This work lays the foundation for both critique and creation: it provides researchers with additional tools to study text systematically, and gives designers a vocabulary for using it more purposefully.

\section*{Supplemental Materials}
\label{sec:supplemental_materials}

All supplemental materials are available on OSF (\href{https://osf.io/swqfc/}{link}), released under a CC BY 4.0 license.
In particular, they include (1) data and analysis files, (2) images from the corpus analyzed, (3) figures used in this paper.

\section*{Figure Credits}
\label{sec:figure_credits}

\textbf{A}. \Cref{fig:teaser} image credits: Img 131: Alyson Hurt / NPR [original~\cite{img131_original}] and Flowing Data [reprint~\cite{img131}], September 2015 \textbf{;} Img 70: Graphic Detail / Economist, October 2011~\cite{img70} \textbf{;} Img 448: Neil King Jr. / Wall Street Journal, September 2012~\cite{img448}

\noindent\textbf{B}. \Cref{fig:functions} image credit (Img 241):
Jonathan House, David Román / Wall Street Journal, December 2011~\cite{img241}

\noindent\textbf{C}. \Cref{fig:factor_loadings} image credits: Img 55: Graphic Detail / Economist, September 2012~\cite{img55} \textbf{;} Img 185: Dion Nissenbaum / Wall Street Journal, June 2012~\cite{img185} \textbf{;} Img 440: Samuel Granados / Washington Post, February 2017~\cite{img440} \textbf{;} Img 403: Jen Christiansen / 
Flowing Data [reprint~\cite{img403}], September 2019 \textbf{;} Img 19: Omri Wallach / \href{https://www.visualcapitalist.com/visualizing-biggest-tech-mergers-and-acquisitions-of-2020/}{Visual Capitalist}, December 2020~\cite{img19} \textbf{;} Img 367: Samuel Velasco / Quanta Magazine, November 2020~\cite{img367} \textbf{;} Img 235: Josh Mitchell and Rachel Louise Ensign / Wall Street Journal, September 2012~\cite{img235} \textbf{;} Img 138: Nadieh Bremer / 
Visual Cinnamon [reprint~\cite{img138}], December 2017


\acknowledgments{%
    The authors wish to thank Michael Correll, Will Wang, Md Dilshadur Rahman, and Simone Laszuk for early comments on the manuscript, as well as the anonymous reviewers for helpful feedback which improved the paper.
    This work was supported in part by the National Science Foundation Graduate Research Fellowship under Grant No. DGE 2146752.
}

\bibliographystyle{abbrv-doi-hyperref-narrow}

\bibliography{0_bib}



\end{document}